\documentclass{article}
\usepackage{amsmath,amssymb,amsfonts}
\usepackage{graphicx}
\usepackage{hyperref}  

\newcommand{\Mvariable}[1]{}

 \providecommand{\imag}[1]{\,i\,}
 \providecommand{\tanbloopfactor}{}

\begin{document}

\author{Graham G. Ross \thanks{g.ross@physics.ox.ac.uk}
and Mario Serna \thanks{serna@physics.ox.ac.uk} \\
\small
Rudolf Peierls Centre for Theoretical Physics,
University of Oxford, 1 Keble Road, Oxford, OX1 3NP  \normalsize}
\title{Unification and fermion mass structure.}
\maketitle

\begin{abstract}
Grand Unified Theories predict relationships between the GUT-scale quark and
lepton masses. Using new data in the context of the MSSM, we update the
values and uncertainties of the masses and mixing angles for the three
generations at the GUT scale. We also update fits to hierarchical patterns
in the GUT-scale Yukawa matrices. The new data shows not all the classic
GUT-scale mass relationships remain in quantitative agreement at small to
moderate $\tan \beta $. However, at large $\tan \beta $, these discrepancies
can be eliminated by finite, $\tan \beta $-enhanced, radiative, threshold
corrections if the gluino mass has the opposite sign to the wino mass.
\end{abstract}

Explaining the origin of fermion masses and
mixings remains one of the most important
goals in our attempts to go beyond the
Standard Model. In this, one very promising
possibility is that there is an underlying
stage of unification relating the couplings
responsible for the fermion masses. However
we are hindered by the fact that the measured
masses and mixings do not directly give the
structure of the underlying Lagrangian both
because the data is insufficient
unambiguously to reconstruct the full fermion
mass matrices and because radiative
corrections can obscure the underlying
structure. In this letter we will address
both these points in the context of the MSSM.

We first present an analysis of the measured mass and mixing angles
continued to the GUT scale. The analysis updates previous work, using the
precise measurements of fermion masses and mixing angles from the
b-factories and the updated top-quark mass from CDF and D0. The resulting
data at the GUT scale allows us to look for underlying patterns which may
suggest a unified origin. We also explore the sensitivity of these patterns
to $\tan \beta $-enhanced, radiative threshold corrections.

We next proceed to extract the underlying Yukawa coupling matrices for the
quarks and leptons. There are two difficulties in this. The first is that
the data cannot, without some assumptions, determine all elements of these
matrices. The second is that the Yukawa coupling matrices are basis
dependent. We choose to work in a basis in which the mass matrices are
hierarchical in structure with the off-diagonal elements small relative to
the appropriate combinations of on-diagonal matrix elements. This is the
basis we think is most likely to display the structure of the underlying
theory, for example that of a spontaneously broken family symmetry in which
the hierarchical structure is ordered by the (small) order parameter
breaking the symmetry. With this structure to leading order the observed
masses and mixing angles determine the mass matrix elements on and above the
diagonal, and our analysis determines these entries, again allowing for
significant $\tan \beta $ enhanced radiative corrections. The resulting form
of the mass matrices provides the \textquotedblleft data\textquotedblright\
for developing models of fermion masses such as those based on a broken
family symmetry.
\begin{table}[tb]
\centerline{
\begin{tabular}{|c|c|p{1.8in}|}
 \hline
 Low-Energy Parameter &
  Value(Uncertainty in last digit(s)) &
  Notes and Reference \cr
  \hline
 $m_u(\mu_L)/m_d(\mu_L)$
   &   0.45(15)
   & PDB  Estimation \cite{PDBook2006} \cr
 $m_s(\mu_L)/m_d(\mu_L)$
   & 19.5(1.5)
   & PDB  Estimation \cite{PDBook2006} \cr
 $m_u(\mu_L)+ m_d(\mu_L)$
 & $\left[ 8.8(3.0),\ 7.6(1.6) \right]$ MeV  &
 PDB, Quark Masses, pg 15 \cite{PDBook2006}. (
 Non-lattice, Lattice )
 \cr
 $Q = \sqrt{\frac{m_s^2 - (m_d + m_u)^2/4}{m_d^2 - m_u^2}}$
   & 22.8(4)
    & Martemyanov and Sopov
\cite{Martemyanov:2005bt} \cr
 $m_s(\mu_L)$
 & $\left[ 103 (20)\, , 95 (20) \right]$ MeV & PDB, Quark Masses, pg 15
\cite{PDBook2006}. [Non-lattice, lattice] \cr
   $m_u(\mu_L)$ & 3(1)\,   MeV & PDB, Quark
Masses, pg 15 \cite{PDBook2006}.
Non-lattice.\cr
 $m_d(\mu_L)$
   & 6.0(1.5)\, MeV &
    PDB, Quark Masses, pg 15 \cite{PDBook2006}. Non-lattice.  \cr
  $m_c(m_c)$ & 1.24(09) GeV & PDB, Quark Masses, pg 16
  \cite{PDBook2006}. Non-lattice.\cr
 $m_b(m_b)$ &
4.20(07) \, GeV & PDB, Quark Masses, pg 16,19
\cite{PDBook2006}. Non-lattice. \cr
 $M_{t}$ & 170.9 (1.9)\,{\rm{GeV}} &  CDF \& D0 \cite{Group:2007bx} Pole Mass   \cr
  $(M_e,M_\mu,M_\tau)$ &
($0.511(15)$,$\ 105.6(3.1)$,$\ 1777(53)$ )
MeV & $3\%$ uncertainty from neglecting $Y^e$
thresholds.  \cr $A$ Wolfenstein parameter &
0.818(17)  &  PDB Ch 11 Eq.~11.25
\cite{PDBook2006}  \cr $\overline{\rho}$
Wolfenstein parameter & 0.221(64)   &  PDB Ch
11 Eq.~11.25 \cite{PDBook2006}  \cr $\lambda$
Wolfenstein parameter &   0.2272(10)     &
PDB Ch 11 Eq.~11.25 \cite{PDBook2006}  \cr
$\overline{\eta}$ Wolfenstein parameter &
0.340(45) &  PDB Ch 11 Eq.~11.25
\cite{PDBook2006}  \cr $|V_{CKM}|$ & $\left(
\begin{matrix} 0.97383(24) & 0.2272(10) & 0.00396(09) \cr
0.2271(10) & 0.97296(24) & 0.04221(80) \cr
0.00814(64) & 0.04161(78) & 0.999100(34)
\end{matrix} \right)$ &   PDB Ch 11 Eq.~11.26
\cite{PDBook2006}  \cr $\sin 2\beta$ from CKM
& 0.687(32) & PDB Ch 11 Eq.~11.19
\cite{PDBook2006} \cr Jarlskog Invariant &
 $3.08(18) \times 10^{-5}$ &  PDB Ch 11
 Eq.~11.26 \cite{PDBook2006}\cr
\hline
 $v_{Higgs}(M_Z)$ & $246.221(20)$ GeV &
 Uncertainty expanded. \cite{PDBook2006}  \cr
 ( $\alpha_{EM}^{-1}(M_Z)$, $\alpha_s(M_Z)$,
 $\sin^2 \theta_W(M_Z)$ ) &  ($\ 127.904(19)$, $\ 0.1216(17)$, $\ 0.23122(15)$) & PDB Sec 10.6
 \cite{PDBook2006} \cr
\hline
\end{tabular}}
\caption{Low-energy observables. Masses in lower-case $m$ are $\overline{MS}$
running masses. Capital $M$ indicates pole mass. The light quark's ($u$,$d$,$%
s$) mass are specified at a scale $\protect\mu _{L}=2\ \mathrm{GeV}$. $%
V_{CKM}$ are the Standard Model's best fit values. }
\label{TableParametersToFitTo}
\end{table}

The data set used is summarized in Table
\ref{TableParametersToFitTo}. Since the fit
of reference \cite{Roberts:2001zy} (RRRV) to
the Yukawa texture was done, the measurement
of the Standard-Model parameters has improved
considerably. We highlight a few of the
changes in the data since 2000: The top-quark
mass has gone from $M_t=174.3 \pm 5$ GeV to
$M_t=170.9 \pm 1.9$ GeV.
In 2000 the Particle Data Book reported $m_{b}(m_{b})=4.2\pm 0.2$ GeV \cite%
{PDBook2000} which has improved to $m_{b}(m_{b})=4.2\pm 0.07$ GeV today. In
addition each higher order QCD correction pushes down the value of $%
m_{b}(M_{Z})$ at the scale of the $Z$ bosons mass. In 1998 $%
m_{b}(M_{Z})=3.0\pm 0.2$ GeV \cite{Fusaoka:1998vc} and today it is $%
m_{b}(M_{Z})=2.87\pm 0.06$ GeV \cite{Baer:2002ek}. The most significant
shift in the data relevant to the RRRV fit is a downward revision to the
strange-quark mass at the scale $\mu _{L}=2$ GeV from $m_{s}(\mu
_{L})\approx 120\pm 50$ MeV \cite{PDBook2000} to today's value $m_{s}(\mu
_{L})=103\pm 20$ MeV. We also know the CKM unitarity triangle parameters
better today than six years ago. For example, in 2000 the Particle Data book
reported $\sin 2\beta =0.79\pm 0.4$ \cite{PDBook2000} which is improved to $%
\sin 2\beta =0.69\pm 0.032$ in 2006 \cite{PDBook2006}. The $\sin 2\beta $
value is about $1.2\,\sigma $ off from a global fit to all the CKM data \cite%
{Lubicz:2007zv}, our fits generally lock onto the global-fit data and
exhibit a $1\,\sigma$ tension for $\sin 2\beta$. Together, the improved CKM
matrix observations add stronger constraints to the textures compared to
data from several years ago.

We first consider the determination of the fundamental mass parameters at
the GUT scale in order simply to compare to GUT predictions. The starting
point for the light-quark masses at low scale is given by the $\chi ^{2}$
fit to the data of Table \ref{TableParametersToFitTo}
\begin{equation}
m_{u}(\mu _{L})=2.7\pm 0.5\ \mathrm{MeV}\ \ m_{d}(\mu _{L})=5.3\pm 0.5\
\mathrm{MeV}\ \ m_{s}(\mu _{L})=103\pm 12\ \mathrm{MeV}.
\label{EqLightQuarkMassConsistent}
\end{equation}%
Using these as input we determine the values
of the mass parameters at the GUT scale for
various choices of $\tan \beta$ but not
including possible $\tan \beta $ enhanced
threshold corrections.  We do this using
numerical solutions to the RG equations. The
one-loop and two-loop RG equations for the
gauge couplings and the Yukawa couplings in
the Standard Model and in
the MSSM that we use in this study come from a number of sources \cite%
{Fusaoka:1998vc} \cite{Chankowski:2001mx}\cite{Ramond:1999vh} \cite%
{Barger:1992ac}. The results are given in the
first five columns of Table \ref{table2}.
These can readily be compared to expectations
in various Grand Unified models. The classic
prediction of $SU(5)$ with third generation
down-quark and charged-lepton masses given by
the coupling
$B\;\overline{5}_{f}.10_{f}.5_{H}$
\footnote{$\overline{5}_{f}$, $10_{f}$ refer
to the $SU(5)$ representations making up a
family of quarks and leptons while $5_{H}$ is
a five dimensional representation of Higgs
scalars.} is $m_{b}(M_{X})/m_{\tau}(M_{X})=1$
\cite{Buras:1977yy}.  This ratio is given in
Table \ref{table2} where it may be seen that
the value agrees at a special low $\tan
\beta$ value but for large $\tan \beta $ it
is some $25\%$ smaller than the GUT
prediction\footnote{We'd like to thank Ilja
Dorsner for pointing out that the $\tan
\beta$ dependence of $m_{b}/m_{\tau}(M_X)$ is
more flat than in previous studies (e.g.
ref.~\cite{Barr:2002mw}).  This change is
mostly due to the higher effective SUSY scale
$M_S$, the higher value of $\alpha_s(M_Z)$
found in global standard model fits, and
smaller top-quark mass $M_t$.}. A similar
relation between the strange quark and the
muon is untenable and to describe the masses
consistently in $SU(5)$ Georgi and Jarlskog
\cite{Georgi:1979df} proposed that the second
generation masses should come instead from
the coupling
$C\;\overline{5}_{f}.10_{f}.45_{H}$ leading
instead to the relation 3$m_{s}(M_{X})/m_{\mu
}(M_{X})=1.$ As may be seen from Table 2 in
all cases this ratio is approximately
$0.69(8)$. The prediction of Georgi and
Jarlskog
for the lightest generation masses follows from the relation $%
Det(M^{d})/Det(M^{l})=1$. This results from the form of their mass matrix
which is given by\footnote{%
The remaining mass matrix elements may be non-zero provided they do not
contribute significantly to the deteminant}
\begin{equation}
M^{d}=\left(
\begin{array}{ccc}
0 & A^{\prime } &  \\
A & C &  \\
&  & B%
\end{array}%
\right) ,\;M^{l}=\left(
\begin{array}{ccc}
0 & A^{\prime } &  \\
A & -3C &  \\
&  & B%
\end{array}%
\right)   \label{GeorgiJarlskog}
\end{equation}%
in which there is a $(1,1)$ texture zero\footnote{%
Below we discuss an independent reason for having a $(1,1)$ texture zero.}
and the determinant is given by the product of the $(3,3)$, $(1,2)$ and
$(2,1)$ elements. If the $(1,2)$ and $(2,1)$ elements are also given by $%
\overline{5}_{f}.10_{f}.5_{H}$ couplings they will be the same in the
down-quark and charged-lepton mass matrices giving rise to the equality of
the determinants. The form of eq(\ref{GeorgiJarlskog}) may be arranged by
imposing additional continuous or discrete symmetries. One may see from
Table \ref{table2} that the actual value of the ratio of the determinants is
quite far from unity disagreeing with the Georgi Jarlskog relation.

In summary the latest data on fermion masses,
while qualitatively in agreement with the
simple GUT relations, has significant
quantitative discrepancies. However the
analysis has not, so far, included the SUSY
threshold corrections which substantially
affect the GUT mass relations at large $\tan
\beta$ \cite{Diaz-Cruz:2000mn}.
\begin{table}[tbh]
\centerline{
\begin{tabular}{|c|c|c|c|c||c|c|}
 \hline
 Parameters & \multicolumn{6}{|c||}{Input SUSY Parameters }  \\
 \hline
 $\tan \beta$ & $1.3$ & $10$ & $ 38$ &  $ 50$ & $38$ & $38$ \cr
 $\gamma_b$ & $0$ & $0$ & $ 0$ &  $0$ & $-0.22$ & $+0.22$ \cr
 $\gamma_d$ & $0$ & $0$ & $ 0$ &  $0$ & $-0.21$ & $+0.21$ \cr
 $\gamma_t$ & $0$ & $0$ & $ 0$ &  $0$ & $0$ &  $-0.44$ \cr
 \hline
 Parameters &
 \multicolumn{6}{|c|}{Corresponding GUT-Scale Parameters
 with
 Propagated Uncertainty}  \cr
 \hline
 $y^t(M_X)$ & $6^{+1}_{-5}$ & $0.48(2)$ & $0.49(2)$ & $0.51(3)$               	 & $0.51(2)$ & $0.51(2)$ \cr
  $y^b(M_X)$ & $0.0113^{+0.0002}_{-0.01}$ & $0.051(2)$ & $0.23(1)$ & $0.37(2)$ 	 & $0.34(3)$ & $0.34(3)$ \cr
  $y^\tau(M_X)$ & $0.0114(3)$ &  $0.070(3)$  & $0.32(2)$ & $0.51(4)$          	 & $0.34(2)$ & $0.34(2)$   \cr
  $(m_u/m_c)(M_X)$ & $0.0027(6)$ & $0.0027(6)$ & $0.0027(6)$ & $0.0027(6)$     	 & $0.0026(6)$ & $0.0026(6)$ \cr
  $(m_d/m_s)(M_X)$ & $0.051(7)$ & $0.051(7)$ & $0.051(7)$ & $0.051(7)$        	 & $0.051(7)$  & $0.051(7)$   \cr
  $(m_e/m_\mu)(M_X)$ & $0.0048(2)$ & $0.0048(2)$ & $0.0048(2)$ & $0.0048(2)$  	 & $0.0048(2)$ & $0.0048(2)$ \cr
  $(m_c/m_t)(M_X)$ & $0.0009^{+0.001}_{-0.00006}$  & $0.0025(2)$ & $0.0024(2)$ & $0.0023(2)$  	& $0.0023(2)$ & $0.0023(2)$ \cr
  $(m_s/m_b)(M_X)$  & $0.014(4)$ & $0.019(2)$ & $0.017(2)$ & $0.016(2)$       	 & $0.018(2)$ & $0.010(2)$ \cr
  $(m_\mu / m_\tau)(M_X)$ & $0.059(2)$  & $0.059(2)$ & $0.054(2)$ & $0.050(2)$      	 & $0.054(2)$ & $0.054(2)$ \cr
  $A(M_X)$ & $0.56^{+0.34}_{-0.01}$ & $0.77(2)$ & $0.75(2)$ & $0.72(2)$       	 & $0.73(3)$  & $0.46(3)$ \cr
  $\lambda(M_X)$ & $0.227(1)$ & $0.227(1)$ & $0.227(1)$ & $0.227(1)$          	 & $0.227(1)$ & $0.227(1)$ \cr
  $\bar{\rho}(M_X)$ & $0.22(6)$ & $0.22(6)$ & $0.22(6)$ &  $0.22(6)$          	 & $0.22(6)$ & $0.22(6)$  \cr
  $\bar{\eta}(M_X)$ & $0.33(4)$ & $0.33(4)$ & $0.33(4)$ & $0.33(4)$           	 & $0.33(4)$  & $0.33(4)$ \cr
  $J(M_X)\,  \times 10^{-5} $ & $1.4^{+2.2}_{-0.2}$  & $2.6(4)$ &  $2.5(4)$ & $2.3(4)$  	 & $2.3(4)$ & $1.0(2)$ \cr
  \hline
  Parameters &
 \multicolumn{6}{|c|}{Comparison with GUT Mass Ratios}  \cr
  \hline
  $(m_b/m_\tau)(M_X)$
     & $1.00^{+0.04}_{-0.4}$
     & $0.73(3)$
     & $0.73(3)$
     & $0.73(4)$       	 & $1.00(4)$ & $1.00(4)$ \cr
  $({3 m_s / m_\mu})(M_X)$
       & $0.70^{+0.8}_{-0.05}$
       & $0.69(8)$
       & $0.69(8)$
       & $0.69(8)$     	 & $0.9(1)$ & $0.6(1)$ \cr
  $({m_d / 3\,m_e})(M_X)$
      & $0.82(7)$
      & $0.83(7)$
      & $0.83(7)$
      & $0.83(7)$      	 & $1.05(8)$ & $0.68(6)$ \cr
  $(\frac{\det Y^d}{\det Y^e})(M_X)$
       & $0.57^{+0.08}_{-0.26}$
       & $0.42(7)$
       & $0.42(7)$
       & $0.42(7)$    	 & $0.92(14)$ & $0.39(7)$ \cr
  \hline
 \end{tabular}}
\caption{The mass parameters continued to the GUT-scale $M_X$ for various
values of $\tan \protect\beta $ and threshold corrections $\protect\gamma %
_{t,b,d}$. These are calculated with the 2-loop gauge coupling and 2-loop
Yukawa coupling RG equations assuming an effective SUSY scale $M_{S}=500$
GeV.}
\label{table2}
\end{table}
A catalog of the full SUSY threshold
corrections is given in \cite{Pierce:1996zz}.
The particular finite SUSY thresholds
discussed in this letter do not decouple as
the super partners become massive. We follow
the approximation described in Blazek, Raby,
and Pokorski (BRP) for threshold corrections
to the CKM elements and down-like mass
eigenstates \cite{Blazek:1995nv}. The finite
threshold corrections to $Y^{e}$ and $Y^{u}$
and are generally about 3\% or smaller
\begin{equation}
\delta Y^{u},\ \delta Y^{d}\lesssim 0.03
\end{equation}%
and will be neglected in our study. The logarithmic threshold corrections
are approximated by using the Standard-Model RG equations from $M_Z$ to an
effective SUSY scale $M_S$.

The finite, $\tan \beta$-enhanced $Y^{d}$ SUSY threshold corrections are
dominated by the a sbottom-gluino loop, a stop-higgsino loop, and a
stop-chargino loop. Integrating out the SUSY particles at a scale $M_{S}$
leaves the matching condition at that scale for the Standard-Model Yukawa
couplings:
\begin{eqnarray}
\delta m_{sch}\,Y^{u\,SM} &=&\sin \beta \ \,Y^{u} \\
\delta m_{sch}\,Y^{d\,SM} &=&\cos \beta \ \,U_{L}^{d\dag }\,\left( 1+%
\tanbloopfactor {\Gamma}^{d}+\tanbloopfactor V_{CKM}^{\dag }\,{\Gamma }%
^{u}\,V_{CKM}\right) \,Y^{d}_{\mathrm{diag}}\,U_{R}^{d} \\
Y^{e\,SM} &=&\cos \beta \,\ Y^{e}.
\end{eqnarray}%
All the parameters on the right-hand side
take on their MSSM values in the
$\overline{DR}$ scheme. The factor $\delta
m_{sch}$ converts the quark running masses
from $\overline{MS}$ to $\overline{DR}$
scheme. The $\beta$ corresponds to the ratio
of the two Higgs VEVs $v_u / v_d=\tan \beta$.
The $U$ matrices decompose the MSSM Yukawa
couplings at the scale $M_{S}$:
$Y^{u}=U_{L}^{u\dag
}Y_{\mathrm{diag}}^{u}U_{R}^{u}$ and
$Y^{d}=U_{L}^{d\dag
}Y_{\mathrm{diag}}^{d}U_{R}^{d}$. The
matrices $Y_{\mathrm{diag}}^{u}$ and
$Y_{\mathrm{diag}}^{d}$ are diagonal and
correspond to the mass eigenstates divided by
the appropriate VEV at the scale $M_{S}$. The
CKM matrix is given by
$V_{CKM}=U_{L}^{u}U_{L}^{d\dag }$. The
left-hand side involves the Standard-Model
Yukawa couplings. The matrices $\Gamma ^{u}$
and $\Gamma ^{d}$ encode the SUSY threshold
corrections.

If the squarks are diagonalized in flavor space by the same rotations that
diagonalize the quarks, the matrices $\Gamma ^{u}$ and $\Gamma ^{d}$ are
diagonal: $\Gamma ^{d}=\mathrm{diag}(\gamma _{d},\gamma _{d},\gamma _{b}),$ $%
\ \Gamma ^{u}=\mathrm{diag}(\gamma _{u},\gamma _{u},\gamma _{t})$. In
general the squarks are not diagonalized by the same rotations as the quarks
but provided the relative mixing angles are reasonably small the corrections
to flavour conserving masses, which are our primary concern here, will be
second order in these mixing angles. We will assume $\Gamma ^{u}$ and $%
\Gamma ^{d}$ are diagonal in what follows.

Approximations for $\Gamma^{u}$ and $\Gamma
^{d}$ based on the mass
insertion approximation are found in \cite{Carena:1999py}\cite{Carena:2002es}%
\cite{Tobe:2003bc}:
\begin{eqnarray}
\gamma _{t} &\approx &y_{t}^{2}\,\mu \,A^{t}\,\frac{\tan \beta }{16\pi ^{2}}%
I_{3}(m_{\tilde{t}_{1}}^{2},m_{\tilde{t}_{2}}^{2},\mu ^{2})\ \ \sim \ \
y_{t}^{2}\,\frac{\tan \beta }{32\pi ^{2}}\frac{\mu \,A^{t}\,}{m_{\tilde{t}%
}^{2}}  \label{Eqgammat} \\
\gamma _{u} &\approx &-g_{2}^{2}\,M_{2}\,\mu \,\frac{\tan \beta }{16\pi ^{2}}%
I_{3}(m_{\chi _{1}}^{2},m_{\chi _{2}}^{2},m_{\tilde{u}}^{2})\ \ \sim \ \ 0
\label{Eqgammau} \\
\gamma _{b} &\approx &\frac{8}{3}\,g_{3}^{2}\,\frac{\tan \beta }{16\pi ^{2}}%
\,M_{3}\,\mu \,I_{3}(m_{\tilde{b}_{1}}^{2},m_{\tilde{b}_{2}}^{2},{M_{3}}%
^{2})\ \ \sim \ \ \frac{4}{3}\,g_{3}^{2}\,\frac{\tan \beta }{16\pi ^{2}}\,%
\frac{\mu \,M_{3}}{m_{\tilde{b}}^{2}}  \label{Eqgammab} \\
\gamma _{d} &\approx &\frac{8}{3}g_{3}^{2}\frac{\tan \beta }{16\pi ^{2}}%
M_{3}\,\mu \,I_{3}(m_{\tilde{d}_{1}}^{2},m_{\tilde{d}_{2}}^{2},{M_{3}}^{2})\
\ \sim \ \ \frac{4}{3}\,g_{3}^{2}\,\frac{\tan \beta }{16\pi ^{2}}\,\frac{\mu
\,M_{3}}{m_{\tilde{d}}^{2}}  \label{Eqgammad}
\end{eqnarray}%
where $I_{3}$ is given by
\begin{equation}
I_{3}(a^{2},b^{2},c^{2})=\frac{a^{2}b^{2}\log \frac{a^{2}}{b^{2}}%
+b^{2}c^{2}\log \frac{b^{2}}{c^{2}}+c^{2}a^{2}\log \frac{c^{2}}{a^{2}}}{%
(a^{2}-b^{2})(b^{2}-c^{2})(a^{2}-c^{2})}.
\end{equation}%
In these expressions $\tilde{q}$ refers to
superpartner of $q$. $\chi ^{j}$ indicate
chargino mass eigenstates. $\mu $ is the
coefficient to the $H^{u}$ $H^{d}$
interaction in the superpotential.
$M_{1},M_{2},M_{3}$ are the gaugino soft
breaking terms. $A^{t}$ refers to the soft
top-quark trilinear coupling. The mass
insertion approximation breaks down if there
is large mixing between the mass eigenstates
of the stop or the sbottom. The right-most
expressions in
eqs(\ref{Eqgammat},\ref{Eqgammab},\ref{Eqgammad})
assume the relevant squark mass eigenstates
are nearly degenerate and heavier than
$M_{3}$ and $\mu$.
  These
 expressions ( eqs
 \ref{Eqgammat} - \ref{Eqgammad}) provide an
 approximate mapping from a supersymmetric
 spectra to the $\gamma_i$ parameters through which we parameterize
 the threshold corrections; however, with the
 exception of Column A of Table \ref{Table4},
 we do not
 specify a SUSY spectra but directly
 parameterize the thresholds corrections through
 $\gamma_i$.

The separation between $\gamma_b$ and
$\gamma_d$ is set by the lack of degeneracy
of the down-like squarks. If the squark
masses for the first two generations are not
degenerate, then there will be a
corresponding separation between the (1,1)
and (2,2) entries of  $\Gamma^d$ and
$\Gamma^u$. If the sparticle spectra is
designed to have a large $A^t$ and a light
stop, $\gamma_t$ can be enhanced and dominate
over $\gamma_b$. Because the charm Yukawa
coupling is so small, the scharm-higgsino
loop is negligible, and $\gamma_u$ follows
from a chargino squark loop and is also
generally small with values around $0.02$
because of the smaller $g_2$ coupling. In our
work, we approximate $\Gamma^u_{22} \sim
\Gamma^u_{11} \sim 0$. The only substantial
correction to the first and second
generations is given by $\gamma_d$
\cite{Diaz-Cruz:2000mn}.

As described in BRP, the threshold corrections leave $|V_{us}|$ and $%
|V_{ub}/V_{cb}|$ unchanged to a good approximation. Threshold corrections in
$\Gamma ^{u}$ do affect the $V_{ub}$ and $V_{cb}$ at the scale $M_{S}$
giving
\begin{equation}
\frac{V_{ub}^{SM}-V_{ub}^{MSSM}}{V_{ub}^{MSSM}}\backsimeq \frac{%
V_{cb}^{SM}-V_{cb}^{MSSM}}{V_{cb}^{MSSM}}\backsimeq -\left( \gamma
_{t}-\gamma _{u}\right) .
\end{equation}
The threshold corrections for the down-quark
masses are given  approximately by
\begin{eqnarray*}
m_{d} & \backsimeq & m_{d}^{0}\, (1+\gamma _{d}+\gamma _{u})^{-1} \\
m_{s} & \backsimeq &m_{s}^{0}\, (1+\gamma _{d}+\gamma _{u})^{-1} \\
m_{b} & \backsimeq &m_{b}^{0}\, (1+\gamma _{b}+\gamma_{t})^{-1}
\end{eqnarray*}%
where the superscript $0$ denotes the mass without threshold corrections.
Not shown are the nonlinear effects which arise through the RG equations
when the bottom Yukawa coupling is changed by threshold effects. These are
properly included in our final results obtained by numerically solving the
RG equations.

Due to our assumption that the squark masses for the first two generations
are degenerate, the combination of the GUT relations given by $\left( \det
M^{l}/\det M^{d}\right) \left( 3\,m_{s}/m_{\mu }\right) ^{2}\left(
m_{b}/m_{\tau }\right) =1$\ is unaffected up to nonlinear effects. Thus we
cannot simultaneously fit all three GUT relations through the threshold
corrections. A best fit requires the threshold effects given by
\begin{eqnarray}
\gamma _{b}+\gamma _{t} &\approx &-0.22\pm 0.02  \label{threshold} \\
\gamma _{d}+\gamma _{u} &\approx &-0.21\pm 0.02.  \label{threshold2}
\end{eqnarray}%
giving the results shown in the penultimate column of Table \ref{table2},
just consistent with the GUT predictions. The question is whether these
threshold effects are of a reasonable magnitude and, if so, what are the
implications for the SUSY\ spectra which determine the $\gamma _{i}?$ From
eqs(\ref{Eqgammab},\ref{Eqgammad}), at $\tan \beta =38$ we have
\begin{equation*}
\frac{\mu \,M_{3}}{m_{\tilde{b}}^{2}}\sim -0.5,\;\ \ \ \frac{m_{\tilde{b}%
}^{2}}{m_{\tilde{d}}^{2}}\sim 1.0
\end{equation*}

The current observation of the muon's $(g-2)_{\mu }$ is $3.4\,\sigma $ \cite%
{Hagiwara:2006jt} away from the
Standard-Model prediction. If SUSY is to
explain the observed deviation, one needs
$\tan \beta >8$ \cite{Everett:2001tq} and
$\mu M_{2}>0$ \cite{Stockinger:2006zn}. With
this sign we must have $\mu M_{3}$ negative
and the $\widetilde{d},$ $\widetilde{s}$
squarks only lightly split from the
$\widetilde{b}$ squarks. $M_{3}$ negative is
characteristic of anomaly mediated SUSY\
breaking\cite{Randall:1998uk} and is discussed in \cite{Hall:1993gn}\cite{Komine:2001rm}%
\cite{Tobe:2003bc}\cite{Pallis:2003aw}.
 Although we have deduced
$M_3<0$ from the approximate
eqs(\ref{Eqgammab},\ref{Eqgammad}), the the
correlation persists in the near exact
expression found in eq(23) of ref
\cite{Blazek:1995nv}.
Adjusting to different squark splitting can
occur in various schemes\cite{Ramage:2003pf}.
However the squark splitting can readily be
adjusted without spoiling the fit because, up
to nonlinear effects, the solution only
requires the constraints implied by
eq(\ref{threshold}), so we may make $\gamma
_{b}>\gamma _{d}$ and hence make
$m_{\tilde{b}}^{2}<m_{\tilde{d}}^{2}$ by
allowing for a small positive value for
$\gamma _{t}.$ In this case $A^{t}$ must be
positive.

It is of interest also to consider  the
threshold effects in the case that $\mu
M_{3}$ is positive. This is illustrated in
the last column of Table \ref{table2} in
which we have reversed the sign of $\gamma
_{d},$ consistent with positive $\mu M_{3}$ ,
and chosen $\gamma _{b}\simeq \gamma _{d}$ as
is expected for similar down squark masses.
The value of $\gamma _{t}$ is chosen to keep
the equality between $m_{b}$ and $m_{\tau }.$
One may see that the other GUT relations are
not satisfied, being driven further away by
the threshold corrections. Reducing the
magnitude of $\gamma _{b}$ and $\gamma _{d}$
reduces the discrepancy somewhat but still
limited by the deviation found in the
no-threshold case (the fourth column of Table
\ref{table2}).

At $\tan \beta $ near $50$ the non-linear
effects are large and $b-\tau $ unification
requires $\gamma _{b}+\gamma _{t}\sim -0.1$
to $-0.15.$ In this case it is possible to
have $t-b-\tau $ unification of the Yukawa
couplings. For $\mu >0,M_{3}>0$, the
\textquotedblleft Just-so\textquotedblright\
Split-Higgs solution of references
\cite{King:2000vp,Blazek:2001sb,Blazek:2002ta,Auto:2003ys}
can achieve this while satisfying both
$b\rightarrow s\ \gamma $ and $(g-2)_{\mu }$
constraints but only with large $\gamma _{b}$
and $\gamma _{t}$ and a large cancellation in
$\gamma _{b}+\gamma _{t}$. In this case, as
in the example given above, the threshold
corrections drive the masses further from the
mass relations for the first and second
generations because $\mu \,M_{3}>0$. It
\textit{is} possble to have $t-b-\tau $
unification with $\mu \,M_{3}<0$,  satisfying
the $b\rightarrow s\ \gamma $ and $(g-2)_{\mu
}$ constraints in which the GUT predictions
for the first and second generation of quarks
is acceptable. Examples include Non-Universal
Gaugino Mediation \cite{Balazs:2003mm} and
AMSB; both have some very heavy sparticle masses ( $\gtrsim 4$ TeV) \cite%
{Tobe:2003bc}. Minimal AMSB with a light sparticle spectra( $\lesssim 1$
TeV), while satisfying $(g-2)_{\mu }$ and $b\rightarrow s\ \gamma $
constraints, requires $\tan \beta $ less than about $30$ \cite%
{Stockinger:2006zn}.

We turn now to the second part of our study in which we update previous fits
to the Yukawa matrices responsible for quark and lepton masses. As discussed
above we choose to work in a basis in which the mass matrices are
hierarchical with the off-diagonal elements small relative to the
appropriate combinations of on-diagonal matrix elements. This is the basis
we think is most likely to display the structure of the underlying theory,
for example that of a spontaneously broken family symmetry, in which the
hierarchical structure is ordered by the (small) order parameter breaking
the symmetry. With this structure to leading order in the ratio of light to
heavy quarks the observed masses and mixing angles determine the mass matrix
elements on and above the diagonal provided the elements below the diagonal
are not anomalously large. This is the case for matrices that are nearly
symmetrical or for nearly Hermitian as is the case in models based on an $%
SO(10)$ GUT.

\begin{table}[tp]
\centerline{
\begin{tabular}{|c|c|c|c|c|c|c|c|}
\hline
 Parameter & 2001 RRRV & Fit A0 &Fit B0 & Fit A1 & Fit B1  & Fit A2  & Fit B2 \cr
 \hline
 $\tan \beta$
          &   Small
          & $1.3$
          & $1.3$
          & $38$
          & $38$
          & $38$
          & $38$
          \cr
 $a'$
         &  ${\mathcal{O}}(1)$
         &  $0$
         &  $0$
         &  $0$
         &  $0$
         &  $-2.0$
         &  $-2.0$ \cr
 $\epsilon_u$
         & $0.05$
         & $0.030(1)$
         & $0.030(1)$
         & $0.0491(16)$          & $0.0491(15)$          & $0.0493(16)$          & $0.0493(14)$            \cr
 $\epsilon_d$
         & $0.15(1)$
         & $0.117(4)$
         & $0.117(4)$
         & $0.134(7)$
         & $0.134(7)$
         & $0.132(7)$
         & $0.132(7)$ \cr
 $|b'|$
         & $1.0$
         & $1.75(20)$
         & $1.75(21)$
         & $1.05(12)$
         & $1.05(13)$
         & $1.04(12)$
         & $1.04(13)$\cr
 ${\rm{arg}}(b')$
         & $90^o$
         & $+\,93(16)^o$
         & $-\,93(13)^o$
         & $+\,91(16)^o$
         & $-\,91(13)^o$
         & $+\,93(16)^o$
         & $-\,93(13)^o$\cr
 $a$
         & $1.31(14)$
         & $2.05(14)$
         & $2.05(14)$
         & $2.16(23)$
         & $2.16(24)$
         & $1.92(21)$
         & $1.92(22)$ \cr
 $b$
         & $1.50(10)$
         & $1.92(14)$
         & $1.92(15)$
         & $1.66(13)$
         & $1.66(13)$
         & $1.70(13)$
         & $1.70(13)$ \cr
 $|c|$
         & $0.40(2)$
         & $0.85(13)$
         & $2.30(20)$
         & $0.78(15)$          & $2.12(36)$
         & $0.83(17)$
         & $2.19(38)$\cr
 ${\rm{arg}}(c)$
         & ${-\,24(3)^o}$
         & ${-\,39(18)^o}$
         & ${-\,61(14)^o}$
         & ${-\,43(14)^o}$          & ${-\,59(13)^o}$          & ${-\,37(25)^o}$          & ${-\,60(13)^o}$          \cr
 \hline
 \end{tabular}}
\caption{ Results of a $\protect\chi^2$ fit of eqs(\protect\ref{EqYuTexture},%
\protect\ref{EqYdTexture}) to to the data in Table \protect\ref{table2} in
the absence of threshold corrections. We set $a^{\prime }$ as indicated and
set $c^{\prime}=d^{\prime}=d=0$ and $f=f^{\prime}=1$ at fixed values. }
\label{Table3}
\end{table}

For convenience we fit to symmetric Yukawa coupling matrices but, as
stressed above, this is not a critical assumption as the data is insensitive
to the off-diagonal elements below the diagonal and the quality of the fit
is not changed if, for example, we use Hermitian forms. We parameterize a
set of general, symmetric Yukawa matrices as:
\begin{eqnarray}
Y^{u}(M_{X}) &=&y_{33}^{u}\left(
\begin{matrix}
d^{\prime }\epsilon _{u}^{4} & b^{\prime }\,\epsilon _{u}^{3} & c^{\prime
}\,\epsilon _{u}^{3}\cr b^{\prime }\,\epsilon _{u}^{3} & f^{\prime
}\,\epsilon _{u}^{2} & a^{\prime }\,\epsilon _{u}^{2}\cr c^{\prime
}\,\epsilon _{u}^{3} & a^{\prime }\,\epsilon _{u}^{2} & 1%
\end{matrix}%
\right) ,  \label{EqYuTexture} \\
Y^{d}(M_{X}) &=&y_{33}^{d}\left(
\begin{matrix}
d\,\epsilon _{d}^{4} & b\,\epsilon _{d}^{3} & c\,\epsilon _{d}^{3}\cr %
b\,\epsilon _{d}^{3} & f\,\epsilon _{d}^{2} & a\,\epsilon _{d}^{2}\cr %
c\,\epsilon _{d}^{3} & a\,\epsilon _{d}^{2} & 1%
\end{matrix}%
\right) .  \label{EqYdTexture}
\end{eqnarray}%
Although not shown, we always choose lepton Yukawa couplings at $M_{X}$
consistent with the low-energy lepton masses. Notice that the $f$
coefficient and $\epsilon _{d}$ are redundant (likewise in $Y^{u}$). We
include $f$ to be able to discuss the phase of the (2,2) term. We write all
the entries in terms of $\epsilon $ so that our coefficients will be ${%
\mathcal{O}}(1)$. We will always select our best $\epsilon $ parameters such
that $|f|=1$.

RRRV noted that all solutions, to leading order in the small expansion
parameters, only depend on two phases $\phi _{1}$ and $\phi _{2} $ given by
\begin{eqnarray}
\phi _{1} &=&(\phi _{b}^{\prime }-\phi _{f}^{\prime })-(\phi _{b}-\phi _{f})
\\
\phi _{2} &=&(\phi _{c}-\phi _{a})-(\phi _{b}-\phi _{f}).
\end{eqnarray}%
where $\phi_x$ is the phase of parameter $x$.
For this reason it is sufficient to consider
only $b^{\prime}$ and $c$ as complex with all
other parameters real.

As mentioned above the data favours a texture zero in the $(1,1)$ position.
With a symmetric form for the mass matrix for the first two families, this
leads to the phenomenologically successful Gatto Sartori Tonin \cite%
{Gatto:1968ss} relation
\begin{equation}
V_{us}(M_{X})\approx \left\vert b\epsilon _{d}-|b^{\prime }|e^{i\,\phi
_{b^{\prime }}}\epsilon _{u}\right\vert \approx \left\vert \sqrt{(\frac{m_{d}%
}{m_{s}})_{0}}-\sqrt{(\frac{m_{u}}{m_{c}})_{0}}e^{i\,\phi _{1}}\right\vert .
\label{EqVusInTermsOfMasses}
\end{equation}%
This relation gives an excellent fit to $V_{us}$ with $\phi _{1}\approx
\,\pm \,90^{o}$, and to preserve it we take $d,$ $d^{\prime }$ to be zero in
our fits. As discussed above, in $SU(5)$ this texture zero leads to the GUT
relation $Det(M^{d})/Det(M^{l})=1$ which, with threshold corrections, is in
good agreement with experiment. In the case that $c$ is small it was shown
in RRRV that $\phi _{1}$ is to a good approximation the CP violating phase $%
\delta $ in the Wolfenstein parameterization.
A non-zero $c$ is necessary to avoid the
relation $V_{ub}/V_{cb}=\sqrt{m_{u}/m_{c}}$
and with the improvement in the data, it is
now necessary to have $c$ larger than was
found in RRRV \footnote{As shown in
ref.~\cite{Matsuda:2006xa}, it is possible,
in a basis with large off-diagonal entries,
to have an Hermitian pattern with the (1,1)
and (1,3) zero provided one carefully
orchestrates cancelations among $Y^{u}$ and
$Y^{d}$ parameters.  We find this approach
requires a strange-quark mass near its upper
limit.}. As a result the contribution to CP\
violation coming from $\phi _{2}$ is at least
$30\%$. The sign ambiguity in $\phi _{1}$
gives rise to an ambiguity in $c$ with the
positive sign corresponding to the larger
value of $c$ seen in Tables \ref{Table3} and
\ref{Table4}.

\begin{table}[tp]
\centerline{
\begin{tabular}{|c|c|c|c|c|c|}
\hline
 Parameter & A & B & C  & B2 & C2 \cr
 \hline
 $\tan \beta$
          & $30$
          & $38$
          & $38$
          & $38$
          & $38$\cr
 $\gamma_b$ & $0.20$
            & $-0.22$
            & $+0.22$
            & $-0.22$
            & $+0.22$ \cr
 $\gamma_t$ & $-0.03$
            & $0$
            & $-0.44$
            & $0$
            & $-0.44$\cr
 $\gamma_d$ & $0.20$
            & $-0.21$
            & $+0.21$
            & $-0.21$
            & $+0.21$\cr
 $a'$
         &  $0$
         &  $0$
         &  $0$
         &  $-2$
         &  $-2$\cr
 \hline
 $\epsilon_u$
         & $0.0495(17)$
         & $0.0483(16)$
         & $0.0483(18)$
         & $0.0485(17)$
         & $0.0485(18)$\cr
 $\epsilon_d$
         & $0.131(7)$
         & $0.128(7)$
         & $0.102(9)$
         & $0.127(7)$
         & $0.101(9)$ \cr
 $|b'|$
         & $1.04(12)$
         & $1.07(12)$
         & $1.07(11)$
         & $1.05(12)$
         & $1.06(10)$\cr
 ${\rm{arg}}(b')$
         & ${\,90(12)^o}$
         & ${\,91(12)^o}$
         & ${\,93(12)^o}$
         & ${\,95(12)^o}$
         & ${\,95(12)^o}$\cr
 $a$
         & $2.17(24)$
         & $2.27(26)$
         & $2.30(42)$
         & $2.03(24)$
         & $1.89(35)$ \cr
 $b$
         & $1.69(13)$
         & $1.73(13)$
         & $2.21(18)$
         & $1.74(10)$
         & $2.26(20)$\cr
 $|c|$
         & $0.80(16)$
         & $0.86(17)$
         & $1.09(33)$
         & $0.81(17)$
         & $1.10(35)$
         \cr
 ${\rm{arg}}(c)$
         & ${-\,41(18)^o}$
         & ${-\,42(19)^o}$
         & ${-\,41(14)^o}$
         & ${-\,53(10)^o}$
         & ${-\,41(12)^o}$
         \cr
 $ Y^u_{33}$
       & $0.48(2)$
       & $0.51(2)$
       & $0.51(2)$
       & $0.51(2)$
       & $0.51(2)$\cr
 $ Y^d_{33}$
       & $0.15(1)$
       & $0.34(3)$
       & $0.34(3)$
       & $0.34(3)$
       & $0.34(3)$\cr
 $ Y^e_{33}$
       & $0.23(1)$
       & $0.34(2)$
       & $0.34(2)$
       & $0.34(2)$
       & $0.34(2)$ \cr
       \hline
  $(m_b/m_\tau)(M_X)$
     & $0.67(4)$
     & $1.00(4)$
     & $1.00(4)$
     & $1.00(4)$
     & $1.00(4)$ \cr
  $({3 m_s / m_\mu})(M_X)$
       & $0.60(3)$
       & $0.9(1)$
       & $0.6(1)$
       & $0.9(1)$
       & $0.6(1)$  \cr
  $({m_d / 3\,m_e})(M_X)$
      & $0.71(7)$
      & $1.04(8)$
      & $0.68(6)$
      & $1.04(8)$
      & $0.68(6)$ \cr
 $ \left|\frac{\det Y^d(M_X)}{\det Y^e(M_X)} \right|$
       & $0.3(1)$
       & $0.92(14)$
       & $0.4(1)$
       & $0.92(14)$
       & $0.4(1)$\cr
 \hline
 \end{tabular}}
\caption{ A $\protect\chi^2$ fit of eqs(\protect\ref{EqYuTexture},\protect
\ref{EqYdTexture}) including the SUSY threshold effects parameterized by the
specified $\protect\gamma_i$.}
\label{Table4}
\end{table}

Table \ref{Table3} shows results from a $\chi ^{2}$ fit of eqs(\ref%
{EqYuTexture},\ref{EqYdTexture}) to to the
data in Table \ref{table2} in the absence of
threshold corrections. The error, indicated
by the term in brackets, represent the widest
axis of the $1\sigma $ error ellipse in
parameter space. The fits labeled `A' have
phases such that we have the smaller
magnitude solution of $|c|$, and fits labeled
`B' have phases such that we have the larger
magnitude solution of $|c|$. As discussed
above, it is not possible unambiguously to
determine the relative contributions of the
off-diagonal elements of the up and down
Yukawa matrices to the mixing angles. In the
fit A2 and B2 we illustrate the uncertainty
associated with this ambiguity, allowing for
$O(1)$ coefficients $a^{\prime }$. In all the
examples in Table \ref{Table3}, the mass
ratios, and Wolfenstein parameters are
essentially the same as in Table
\ref{table2}.

The effects of the large $\tan \beta $
threshold corrections are shown in Table
\ref{Table4}. The threshold corrections
depend on the details of the SUSY\ spectrum,
and we have displayed the effects
corresponding to a variety of choices for
this spectrum. Column A corresponds to a
\textquotedblleft standard\textquotedblright\
SUGRA fit - the benchmark Snowmass Points and
Slopes (SPS) spectra 1b of
ref(\cite{Allanach:2002nj}). Because the
spectra SPS 1b has large stop and sbottom
squark mixing angles, the approximations
given in eqns(\ref{Eqgammat}-\ref{Eqgammad})
break down, and the value for the correction
$\gamma_i$ in Column A need to be calculated
with the more complete expressions in BRP
\cite{Blazek:1995nv}. In the column A fit and
the next two fits in columns B and C, we set
$a^{\prime }$ and $c^{\prime }$ to zero.
Column B corresponds to the fit given in the
penultimate column of Table \ref{table2}
which agrees very well with the simple GUT
predictions. It is characterized by the
\textquotedblleft
anomaly-like\textquotedblright\ spectrum with
$M_{3}$ negative. Column C examines the
$M_{3}$ positive case while maintaining the
GUT\ prediction for the third generation
$m_{b}=m_{\tau }.$ It corresponds to the
\textquotedblleft Just-so\textquotedblright\
Split-Higgs solution. In the fits A, B and C
the value of the parameter $a$ is
significantly larger than that found in RRRV.
This causes problems for models based on
non-Abelian family symmetries, and it is of
interest to try to reduce $a$ by allowing
$a^{\prime },$ $b^{\prime }$ and $c^{\prime
}$ to vary  while remaining
${\mathcal{O}}(1)$ parameters. Doing this for
the fits B and C leads to the fits B2 and C2
given in Table \ref{Table4} where it may be
seen that the extent to which $a$ can be
reduced is quite limited. Adjusting to this
is a challenge for the broken family-symmetry
models.

Although we have included the finite corrections to match the MSSM theory onto the standard model at an effective SUSY scale $M_S=500$ GeV, we have not included finite corrections from matching onto a specific GUT model.  Precise threshold corrections cannot be rigorously calculated without a specific GUT model.  Here we only estimate the order of magnitude of corrections to the mass relations in Table \ref{table2} from matching the MSSM values onto a GUT model at the GUT scale.  The $\tan \beta$ enhanced corrections in eq(\ref{Eqgammat}-\ref{Eqgammad}) arise from soft SUSY breaking interactions and are suppressed by factors of $M_{SUSY}/M_{GUT}$ in the high-scale matching.  Allowing for ${\mathcal{O}}(1)$ splitting of the mass ratios of the heavy states, one obtains corrections to $y^b/y^\tau$ (likewise for the lighter generations) of ${\mathcal{O}}(\frac{g^2}{(4\pi)^2})$ from the $X$ and $Y$ gauge bosons and $ {\mathcal{O}}(\frac{y_b^2}{(4 \pi)^2})$ from colored Higgs states. Because we have a different Higgs representations for different generations, these threshold correction will be different for correcting the $3 m_s / m_\mu$ relation than the $m_b / m_\tau$ relation.  These factors can be enhanced in the case there are multiple Higgs representation.  For an $SU(5)$ SUSY GUT these corrections are of the order of $2\,\%$.  Plank scale suppressed operators can also induce corrections to both the unification scale \cite{Hill:1983xh} and may have significant effects on the masses of the lighter generations \cite{Ellis:1979fg}.  In the case that the Yukawa texture is given by a broken family symmetry in terms of an expansion parameter $\epsilon$, one expects model dependent corrections of order $\epsilon$ which may be significant.

In summary, in the light of the significant
improvement in the measurement of fermion
mass parameters, we have analyzed the
possibility that the fermion mass structure
results from an underlying supersymmetric GUT
at a very high-scale mirroring the
unification found for the gauge couplings.
Use of the RG equations to continue the mass
parameters to the GUT scale shows that,
although qualitatively in agreement with the
GUT\ predictions coming from simple Higgs
structures, there is a small quantitative
discrepancy. We have shown that these
discrepancies may be eliminated by finite
radiative threshold corrections involving the
supersymmetric partners of the Standard-Model
states. The required magnitude of these
corrections is what is expected at large
$\tan \beta $, and the form needed
corresponds to a supersymmetric spectrum in
which the gluino mass is negative with the
opposite sign to the Wino mass. We have also
performed a fit to the recent data to extract
the underlying Yukawa coupling matrices for
the quarks and leptons. This is done in the
basis in which the mass matrices are
hierarchical in structure with the
off-diagonal elements small relative to the
appropriate combinations of on-diagonal
matrix elements, the basis most likely to be
relevant if the fermion mass structure is due
to a spontaneously broken family symmetry. We
have explored the effect of SUSY\ threshold
corrections for a variety of SUSY\ spectra.
The resulting structure has significant
differences from previous fits, and we hope
will provide the \textquotedblleft
data\textquotedblright\ for developing models
of fermion masses such as those based on a
broken family symmetry.

M.S. acknowledges support from the United States Air Force Institute of
Technology. The views expressed in this letter are those of the authors and
do not reflect the official policy or position of the United States Air
Force, Department of Defense, or the US Government.

\newcommand{\noopsort}[1]{} \newcommand{\printfirst}[2]{#1}
  \newcommand{\singleletter}[1]{#1} \newcommand{\switchargs}[2]{#2#1}
\providecommand{\href}[2]{#2}\begingroup\raggedright\endgroup

\end{document}